\documentclass[prl,twocolumn,superscriptaddress,amsmath,amssymb]{revtex4}

\usepackage{graphicx}
\usepackage{bm}
\usepackage{amsmath}
\usepackage{amssymb}
\usepackage{amsthm}
\usepackage{multirow}
\usepackage{setspace}

\newcommand {\Fig}[1] {Figure~\ref{#1}}
\newcommand {\fig}[1] {Figure~\ref{#1}}   

\newcommand {\eqn}[1] {Equation~(\ref{#1})}
\newcommand {\tab}[1] {Table~\ref{#1}} 

\newcommand{\beq}{\begin{equation}}
\newcommand{\eeq}{\end{equation}}

\newcommand{\natc}{N@C$_{60}$}

\newcommand{\fivenatc}{$^{15}$N@C$_{60}$}

\newcommand{\csixty}{C$_{60}$}

\newcommand{\nfifteen}{$^{15}$N}

\newcommand{\p}{$^{31}$P}

\newcommand{\beqa}{\begin{eqnarray}}
\newcommand{\eeqa}{\end{eqnarray}}

\newcommand{\ket}[1]{\left| #1 \right\rangle}

\newcommand{\JCP}{J. Chem. Phys.}

\newcommand{\JMR}{J. Mag. Res.}

\newcommand{\PR}{Phys. Rev.}

\newcommand{\PRB}{Phys. Rev. B}

\newcommand{\tonee}{$T_{\rm{1e}}$}
\newcommand{\ttwon}{$T_{\rm{2n}}$}
\newcommand{\ttwoe}{$T_{\rm{2e}}$}

\begin{document}

\title{Coherent state transfer between an electron- and nuclear spin in \fivenatc\ }

\author{Richard~M.~Brown}
\email{richard.brown@materials.ox.ac.uk} \affiliation{Department of Materials, Oxford University, Oxford OX1 3PH, UK}

\author{Alexei~M.~Tyryshkin}
\affiliation{Department of Electrical Engineering, Princeton
University, Princeton, NJ 08544, USA}

\author{Kyriakos Porfyrakis}
\affiliation{Department of Materials, Oxford University, Oxford OX1 3PH, UK}

\author{Erik M.~Gauger}
\affiliation{Department of Materials, Oxford University, Oxford OX1 3PH, UK}

\author{Brendon W.~Lovett}
\affiliation{Department of Materials, Oxford University, Oxford OX1 3PH, UK}
\affiliation{School of Engineering and Physical Sciences, Heriot Watt University, Edinburgh EH14 4AS, UK}

\author{Arzhang Ardavan}
\affiliation{CAESR, Clarendon Laboratory, Department of Physics, Oxford University, Oxford OX1 3PU, UK}

\author{S.~A.~Lyon}
\affiliation{Department of Electrical Engineering, Princeton University, Princeton, NJ 08544, USA}

\author{G.~Andrew.~D.~Briggs}
\affiliation{Department of Materials, Oxford University, Oxford OX1 3PH, UK}

\author{John~J.~L.~Morton}
\affiliation{Department of Materials, Oxford University, Oxford OX1 3PH, UK}
\affiliation{CAESR, Clarendon Laboratory, Department of Physics, Oxford University, Oxford OX1 3PU, UK}

\date{\today}
\begin{abstract}
Electron spin qubits in molecular systems offer high reproducibility and the ability to self assemble into larger architectures. However, interactions between neighbouring qubits are `always-on' and although the electron spin coherence times can be several hundred microseconds, these are still much shorter than typical times for nuclear spins. Here we implement an electron-nuclear hybrid scheme which uses coherent transfer between electron and nuclear spin degrees of freedom in order to both controllably turn on/off dipolar interactions between neighbouring spins and benefit from the long nuclear spin decoherence times (\ttwon). We transfer qubit states between the electron and \nfifteen\ nuclear spin in \fivenatc\ with a two-way process fidelity of 88\%, using a series of tuned microwave and radiofrequency pulses and measure a nuclear spin coherence lifetime of over 100~ms.\end{abstract}

\maketitle

Hybrid quantum computing schemes aim to harness the benefits of multiple quantum degrees of freedom through the coherent transfer of quantum information between them. Such transfer has previously been shown between light and atomic ensembles~\cite{julsgaard04, Chan05}, as well as electron to nuclear spin states in nitrogen vacancies~\cite{dutt07} and \p\ donors~\cite{morton2008}, and progress is being made towards coupling electron spin ensembles to superconducting qubits~\cite{sch10, kubo10}. Common motivations for state transfer between electron to nuclear spin qubits include the much longer decoherence times typically exhibited by the nuclear spin, and also the weaker dipolar interaction between nuclear spins which allows interactions between neighbouring qubits to be effectively turned off~\cite{kane, ju07, dutt07, morton2008, yang10}.  Both effects can be attributed to the relatively weak nuclear magnet moment, typically 3 orders of magnitude smaller than an electron spin.
Thus, a powerful hybrid model for quantum computing is one where the electron spin qubit (which is more readily polarised and more quickly manipulated) is used for initialisation and processing, while the nuclear spin is used as a memory. The presence of the electron spin also offers considerable advantages for the readout of a single qubit, either of the electron spin state directly~\cite{morello2010, jelezko04}, or a quantum non-demolition measurement of the nuclear spin~\cite{sarovar, neu10}.

Endohedral fullerenes (atoms held within a carbon cage) offer promise as a molecular qubits due to their exceptionally long electron decoherence times~\cite{brown10, morton06, morton07} and convenient coupling to a local nuclear spin. In particular, \natc\ has been used to demonstrate polarisation transfer from the electron to the nuclear spin and subsequent `bang-bang' decoupling~\cite{bangbang, nc60nuc}, as well as generation of  pseudo-entanglement between the electron and nuclear spin~\cite{mehring04}. The advantages of a molecular system include the ability to produce larger arrays that can be engineered to control electron dipolar interaction~\cite{fullerene06, har02, yang10, guzman}. Molecular approaches offer self assembly of highly reproducible structures~\cite{khl04, khl05}, but are limited by the `always-on' nature of dipolar interactions between neighbouring electron spins. In this Letter we employ a molecular high spin system, comprising an \nfifteen\ atom encapsulated within a carbon cage (\fivenatc). We select a spin concentration such that the electron dipolar coupling is of the order of $\sim$2~kHz. We transfer a coherent state from the electron spin degree of freedom to the nuclear spin, and show that this is able to effectively turn off the dipolar coupling between nearby molecules. We study the fidelity of the transfer process and investigate the decoherence time of the nitrogen nuclear spin at low spin concentrations.
   
\begin{figure}[t] \centerline
{\includegraphics[width=3.5in]{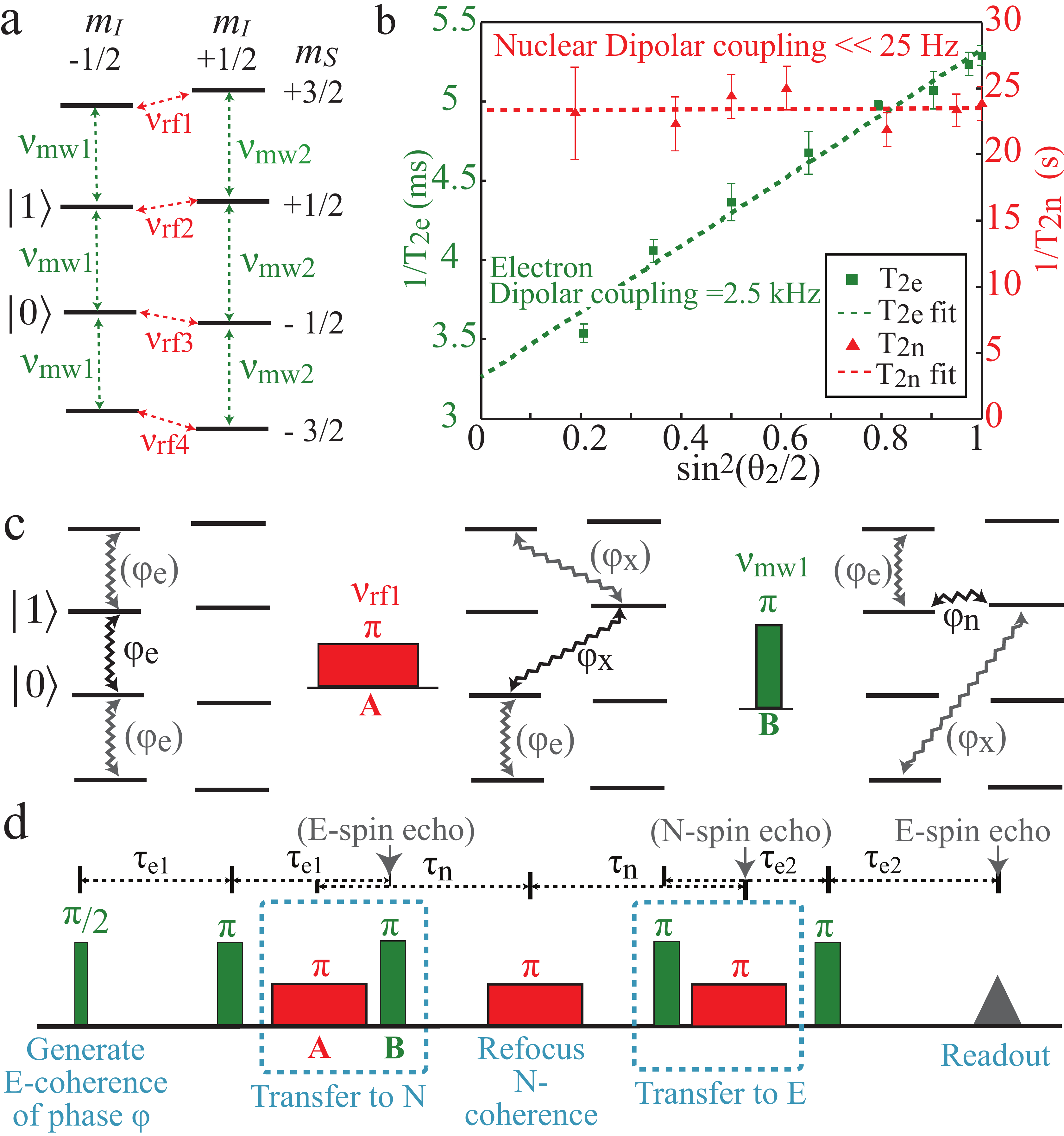}} \caption{(color online). a) The coupled electron spin ($S=3/2$), nuclear spin ($I=1/2$) system for \fivenatc\ leads to 8 levels. A qubit can be represented across an electron spin transition where $m_I=-\frac{1}{2}$, $m_S=\pm\frac{1}{2}$ are denoted states $\ket0$ and $\ket1$. Transitions can be addressed via resonant microwave (mw) and radiofrequency (rf) pulses. b) Varying the length of the refocusing pulse $\theta_{2}$ (see main text) allows a measure of the dipolar coupling between spin qubits, found to be much weaker when they reside in the nuclear spin than in the electron spin (data taken at 20 and 40~K respectively). Due to the limited nuclear spin coherence time, only an upper bound for the nuclear dipolar coupling strength can be extracted. c) Transfer of a qubit state from an electron spin degree of freedom to the \nfifteen\ nuclear spin, within the $m_S=+\frac{1}{2}$ subspace. Coherences are depicted by zig-zag lines and `unwanted' coherences generated by the initial $\pi/2$ pulse on the $S=3/2$ electron spin are shown in (grey). At the end of the transfer sequence, such coherences will decay on the timescale of \ttwoe\ or faster, while the stored qubit will lose coherence on the timescale of \ttwon. d) The full two-way transfer sequence.} \label{transfer} 
\end{figure}

The \fivenatc\ system consists of an $S=3/2$ electron spin coupled via an isotropic hyperfine interaction of 22~MHz to the \nfifteen\ nuclear spin ($I=1/2$). Under an applied magnetic field of $\sim0.35~T$, the energy level diagram is shown in \Fig{transfer}(a)---  this produces a doublet in the electron spin resonance (ESR) spectrum where each line corresponds to a state of $m_I$ \cite{mor07}. To first order, the three electron $\Delta m_S=1$ transitions in each $m_I$ subspace have the same energy and cannot be addressed individually~\cite{mortonESEEM}.
Thus a $\pi/2$ ESR pulse (selective on one $m_I$ state) produces coherences across all three pairs of levels (with $\Delta m_S=1$). For convenience, we will refer to an electron coherence between $m_S$ levels $+\frac{1}{2}:-\frac{1}{2}$ as an \emph{inner} coherence, and those between $m_S$ levels $\pm\frac{3}{2}:\pm\frac{1}{2}$ as \emph{outer} coherences.  A qubit can be represented by the inner pair of $m_S$ levels, in the subspace of $m_I=\frac{1}{2}$ (see \Fig{transfer}(a)). The  \ttwoe\ we report here refers to this \emph{inner} coherence~\cite{supp}.

We used dilute \fivenatc\ in a \csixty\ matrix ($2.5 \times 10^{15}$ spins/cm$^3$), prepared by arc discharge and ion bombardment. The sample was purified using high performance liquid chromotography (HPLC) to remove unwanted amorphous material, placed in a quartz EPR tube and pumped for several hours to remove paramagnetic O$_{2}$ before sealing. For pulsed EPR, we used an X-band (9-10~GHz) Bruker Elexsys spectrometer and a low temperature helium-flow cryostat (Oxford CF935). Typical pulse lengths are 80~ns for a mw $\pi$ pulse using a travelling wave tube (TWT) amplifier and 10~$\mu$s for an rf $\pi$ pulse using a 500W Amplifier Research solid state amplifier. \\

The effect of the dipolar interaction between the electron spins of \fivenatc\ can be observed through a standard Hahn echo experiment ($\pi/2-\tau-\theta_{2}-\tau-echo$) used to measure the electron spin decoherence time (\ttwoe)~\cite{schweiger01}. In this experiment the $\theta_{2}$ pulse (which is typically $\pi$) acts to refocus effects such as magnetic field inhomogenity as well as other interactions experienced by the spin which are constant on the timescale of $\tau$. However, if the $\theta_{2}$ pulse flips both the spin that is observed and a dipolar-coupled neighbouring spin, the effect of this interaction is not refocused and the effective \ttwoe\ is reduced (this effect is termed instantaneous diffusion). If the $\theta_{2}$ pulse is shortened it will act to refocus only a sub-set of spins and mimic a homogeneously dilute spin sample~\cite{schweiger01, klauder62, tyr03}. Plotting 1/\ttwoe\ vs sin$^{2}(\theta_{2}/2)$, \ttwoe\ can be then be extended from 190~$\mu$s using the standard Hahn echo sequence to an extrapolated 300~$\mu$s in the limit $\theta_{2}$=0 (see ~\Fig{transfer}(b) and ~\cite{supp}). From this measurement we extract a dipolar coupling of 2.5~kHz between electron spins at the average \natc\ separation, that we will show is not present between nuclear spins.

To probe the nuclear spin qubit we employ the transfer sequence shown in \Fig{transfer} to propagate an electron coherence to a nuclear coherence. The implementation of this sequence is complicated compared to previous studies~\cite{morton2008} by the presence of the $S=3/2$ electron spin, such that the initial $\pi/2$ mw pulse produces both an \emph{inner} coherence and unwanted \emph{outer} coherences. The application of an rf $\pi$ pulse on the $m_S=+\frac{1}{2}$ transition (a controlled-NOT in quantum gate terminology) then transfers the qubit to an electron-nuclear cross coherence ($\varphi_{x}$). A mw $\pi$ pulse selective on $m_I=-\frac{1}{2}$ then completes the SWAP operation to produce a nuclear coherence ($\varphi_{n}$). Unwanted \emph{outer} coherences generated during the sequence remain as both electron- and multiple-quantum coherences, which decay on the timescale of the electron spin decoherence time (\ttwoe) or faster~\cite{supp}. The desired nuclear spin coherence can then be stored for many milliseconds before transfer back to the electron spin via a reverse of the sequence and readout by a conventional electron spin (Hahn) echo. The full sequence is shown in \Fig{transfer} with the addition of carefully placed pulses to refocus the effect of inhomogeneous broadening on the spin packets in electron, nuclear and multiple quantum coherences. It is not possible to store the qubit within a nuclear coherence in the $m_S=\pm\frac{3}{2}$ subspaces using this sequence, but they are considered in the supplementary material~\cite{supp}.
 
\begin{figure}[t] \centerline
{\includegraphics[width=3.5in]{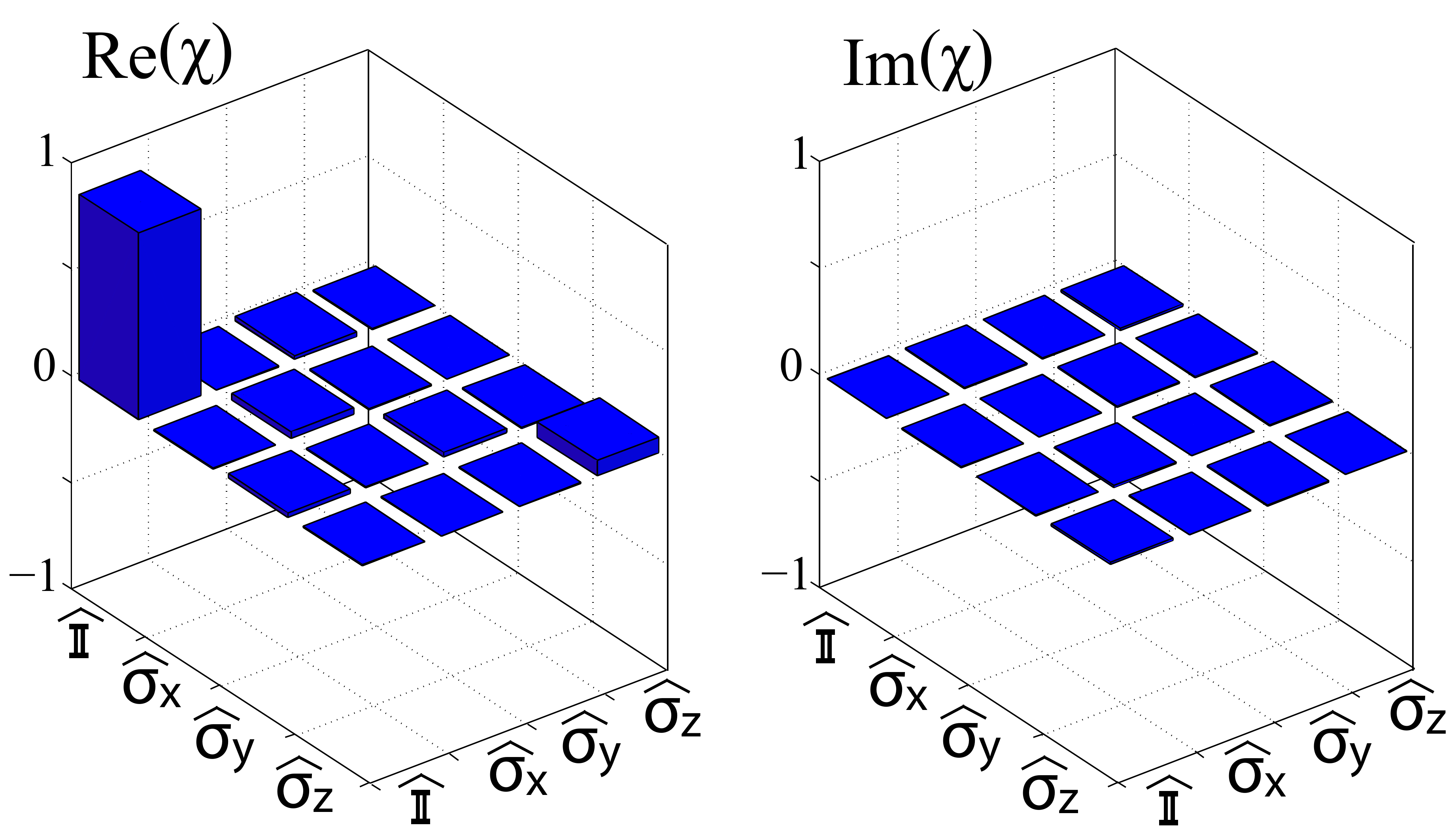}} \caption{(color online). Quantum process tomography matrix ($\chi$) for the transfer of a qubit state from the electron to the nuclear degree of freedom and back, in the basis ($\mathbb{I}$, $\sigma_{x}$, $\sigma_{y}$, $\sigma_{z}$). $\chi$ is evaluated given reference and recovered matrices and gives a fidelity compared to a perfect $\mathbb{I}$ of 0.88. } \label{fidelity}
\end{figure}

There are a number of ways to confirm that the recovered electron spin echo arises solely from a state which was stored in a nuclear spin degree of freedom. One method is to apply a time-varying phase shift to the nuclear spin (e.g., a geometric phase gate~\cite{entanglement2010}) and observe a corresponding phase shift in the electron spin echo. This measurement shows no evidence of any other contribution to the electron spin echo, and is described in more detail in the supplementary material~\cite{supp}. Ultimately, the success of the transfer scheme is shown by the ability to recover any input state with high fidelity after storage in the nuclear spin. This is achieved by exciting the full electronic and nuclear transitions, made possible by the short pulse lengths used and the narrow intrinsic sample ESR and NMR linewidths $<0.6$~MHz and 15~kHz, respectively. We prepare the input states, $\pm$X, $\pm$Y and $\pm$ Z, by varying the phase of the initial $\pi$/2 mw pulse ($\pm$X,$\pm$Y), applying an initial $\pi$ pulse (+Z) or by removing the initial pulse ($-$Z). Using quantum process tomography we can then extract the process matrix for the transfer scheme, $\chi$, in the basis ($\mathbb{I}$, $\sigma_{x}$, $\sigma_{y}$, $\sigma_{z}$)~\cite{niel00PT}. To accurately evaluate $\chi$ we compare the recovered states from the transfer sequence with those given by an ordinary Hahn echo ($\tau=\tau_{e1}+\tau_{e2}$). Thus, $\chi$ incorporates any losses at the storage or retrieval step, as well as during the storage period in the nuclear spin, but not any errors associated with the state generation or measurement. \fig{fidelity} shows the measured $\chi$, giving a fidelity of 0.88, compared to the ideal Identity process ($\mathbb{I}$). We attribute this fidelity primarily due to errors in the transfer pulses---the use of composite mw pulses using the BB1 sequence improves the fidelity to around 94\% and we would expect further improvement with composite rf pulses (see supplementary material~\cite{supp}).

The nuclear decoherence time (\ttwon) can be found by varying the time the qubit is held within the nuclear spin state (2$\tau_{n}$). The resulting exponential decay in echo intensity gives \ttwon\ as long as 135$\pm$13~ms (at 10~K). At this temperature, \ttwoe\ is 160~$\mu$s and thus the nuclear memory gives almost three orders of magnitude improvement in the decoherence time. Nuclear dipolar coupling can be assessed through the effect on \ttwon\ of an `instantaneous diffusion' experiment, similar to that applied on the electron spin. Reducing the length of the nuclear refocusing pulse (0.2 $\le $ sin$^{2}(\theta_{rf}/2) \le$ 1.0, see ~\Fig{transfer}(b)) results in no appreciable change in \ttwon\ at 20~K. Hence, we show the dipolar interaction is $\ll25$ Hz and effectively `turned off' between neighbouring nuclear spins~\footnote{The value of \ttwon\ means we can extract only an upper bound for the dipolar coupling, which is expected to be $\sim10^{6}$ weaker between nuclear spins compared to electron spins and thus of order milliHz.}.

\begin{figure}[t] \centerline
{\includegraphics[width=3.5in]{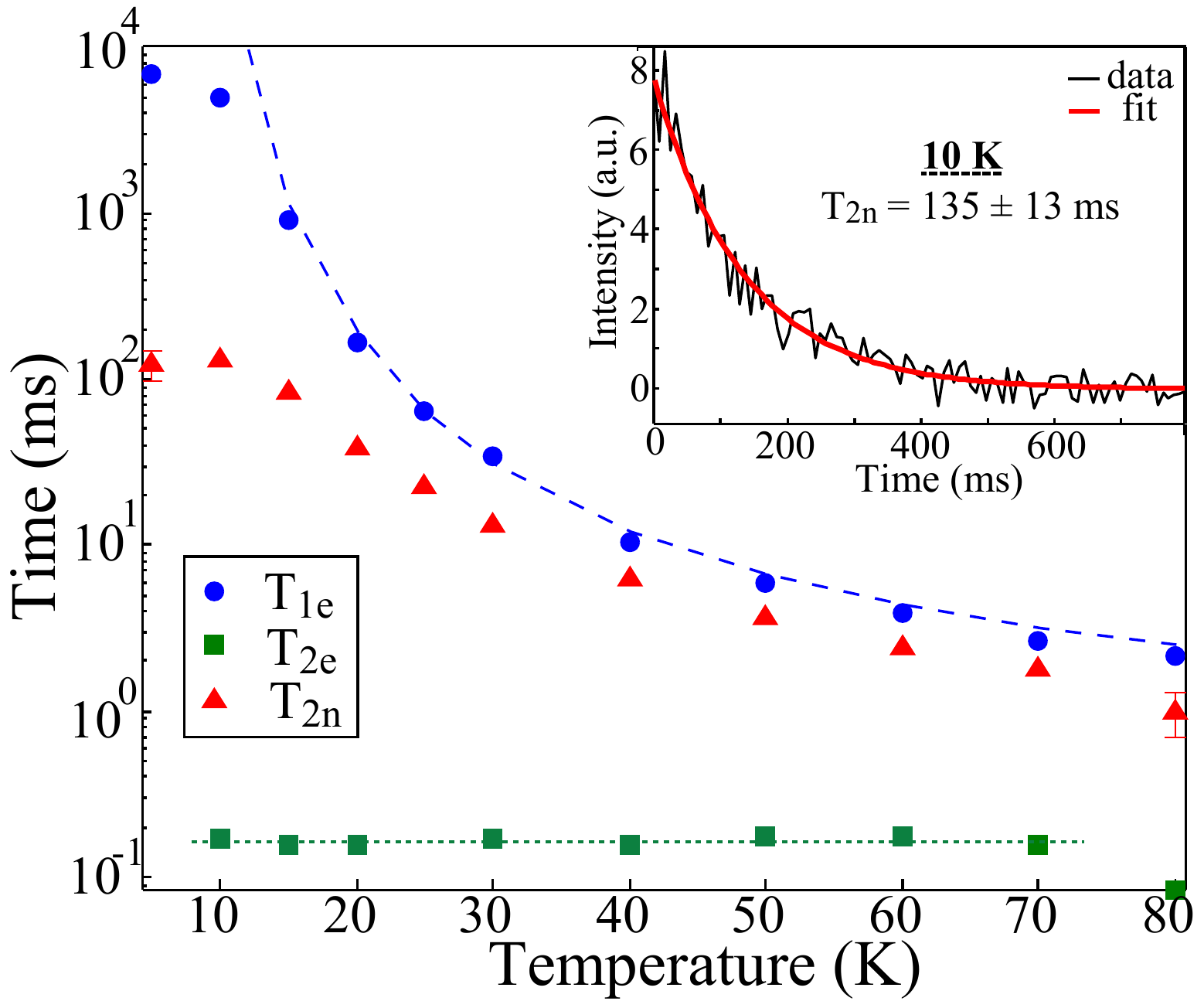}} \caption{{(color online). Relaxation and decoherence times as a function of temperature: \tonee\ (blue, circle), \ttwoe\ (green, square), \ttwon\ (red, triangle), from monoexponential fits with error less than the marker size unless shown. The dashed line is a fit to an Arrhenius temperature dependence for \tonee. The dotted line for \ttwoe\ is a guide. Inset, the nuclear decoherence curve with a monoexponential fit to 135$\pm$13~ms at 10~K.}}
\label{temp}
\end{figure} 

The temperature dependence of the fundamental spin relaxation parameters in the system are shown in \Fig{temp}. The electron relaxation time, \tonee\,  (measured by a standard inversion recovery sequence ($\pi-\tau-\pi/2-T-\pi-T-echo$~\cite{schweiger01})) is shown to increase exponentially with decreasing temperature. This follows an Arrhenius dependence, consistent with a two-phonon process resonant with an excited vibrational mode~\cite{morton06, morton07} and can reach several seconds at low temperatures. Electron spin flips (whose timescale is characterised by \tonee) ultimately act to limit the nuclear coherence time. In the temperature range, 50--80~K, we find that \ttwon\ follows \tonee\ with the experimentally determined relationship, \ttwon~$\sim$~0.6 \tonee. Below 50~K, a secondary mechanism is evident that limits the nuclear decoherence time to $\sim$130~ms.
We analytically model relaxation in the system by applying the Lindblad equation, with the relevant raising and lowering operators, to a given initial state (e.g.\ a pure nuclear coherence, for \ttwon\ or inverted electron state, for \tonee):
\beq
\dot{\rho}=-{\gamma_{a}}(\rho S^{\mp}S^{\pm} +S^{\mp}S^{\pm}\rho -2S^{\pm} S^{\mp})-i[\mathcal{H},\rho]
\label{lindblad}
\eeq
where $\gamma_{a}$ represents both $\gamma_{1}$, the electron spin relaxation rate between the $m_{S}$ levels $\pm\frac{3}{2} \leftrightarrow \pm\frac{1}{2}$ and $\gamma_{2}$, the relaxation rate between $m_{S}$ levels $\frac{1}{2} \leftrightarrow -\frac{1}{2}$. The raising and lowering operators are given by $S^{+}$ and $S^{-}$. Applying relaxation in the high temperature limit and assuming no direct nuclear relaxation, the relevant density matrix elements show a nuclear dephasing rate, $\Gamma_{n}$ = (3$\gamma_{1}$ + 4$\gamma_{2}$). Similarly, taking Eq.~\ref{lindblad} and solving a  series of coupled linear equations the electron polarisation is expressed in terms of two parts: 
\beq
P(t) = \alpha e^{-\lambda_{-}t}+\beta e^{-\lambda_{+}t}
\label{t1e}
\eeq
where $\alpha$ and $\beta$ are prefactors which are a function of $\gamma_{1}$ and $\gamma_{2}$, and the eigenvalues $\lambda$ are given by:
\beq
\lambda_{\pm}=\Gamma_{n}\pm\sqrt{(3\gamma_{1})^{2} + (4\gamma_{2})^{2}}
\label{eigenvalue}
\eeq
It can be shown that the slower decaying component, $\lambda_{-}$, must be dominant, which gives a maximum ratio of $\lambda_{-}=\Gamma_{e}\sim~0.3~\Gamma_{n}$~(\ttwon\ $\sim$~0.3~\tonee), when $3 \gamma_1 = 4 \gamma_2$. 
To reconcile this ratio with the experimentally obtained \ttwon\ $\sim$~0.6~\tonee\ additional relaxation processes can be included in the model, for instance, if $\gamma_{3}$ is given by $m_{S}=\pm\frac{3}{2} \leftrightarrow m_{S}=\mp\frac{1}{2}$ then when $\gamma_{1}=\gamma_{3}\ge\gamma_{2}$ a theoretical \ttwon\ of up to 2/3~\tonee, can be found.

In conclusion, we have reported the coherent transfer of qubit states between electron- and nuclear spin degrees of freedom, in a high spin system. The quantum process tomography of the two-way transfer shows a fidelity of 88$\%$, while we measure a nuclear decoherence time of up to 130~ms, almost three orders of magnitude longer than the electron spin coherence time. Thus, the \nfifteen\ nuclear spin can be employed as both a quantum memory and to controllably turn off dipolar interactions present between electron spin qubits. This is a crucial element in the realisation of fullerene hybrid QIP schemes that exploit the nuclear and electron spin~\cite{fullerene06, ju07, yang10}, especially given recent work in producing larger fullerene architectures~\cite{guzman}. Alternatively, the coupling between spin ensembles and cavities could be exploited~\cite{sch10, kubo10}, along with the storage of multiple microwave excitations~\cite{wu10}, to produce a robust multimode nuclear memory.\\ 

We thank Stephanie Simmons for helpful discussions and instrumentation support. The research is supported by the NSF through the Princeton MRSEC under Grant No. DMR-0213706 and the EPRSC through IMPRESS (EP/D074398/1), and CAESR (EP/D048559/1). J.J.L.M, B.W.L and A.A are supported by the Royal Society.

\onecolumngrid
\appendix
\pagebreak

\section{Supplementary Material}

The supplementary information both supports claims made in the main text and elucidates on further experiments that were conducted. A major part of this document illustrates the intricacies of the spin 3/2 system and the experiments to understand this, with results that are often not intuitive.

\subsection{Electron Coherence: Instantaneous diffusion}
The instantaneous diffusion experiment shows the effect of spin dipolar interaction by varying the length of the refocusing pulse $\theta_{2}$ in a Hahn echo experiment ($\pi/2-\tau-\theta_{2}-\tau-echo$). The relationship between the effective decoherence time ($T_{2e, \rm ID}$) and $\theta_{2}$ for a spin 1/2 system is well known and given by Refs~\cite{sch01id, salikhov81}. In the high spin case a scaling factor, $\kappa$, needs to be applied to account for a higher spin number. We have calculated a scaling factor of 2 for a spin 3/2 system in the high temperature limit, which is confirmed by Walstedt and Walker~\cite{wal74}, yielding:
\beq
1/T_{2e, \rm ID} = \frac{\mu_{0}\pi g^{2}\beta^{2}C\kappa}{9\sqrt{3}\hbar}\sin^{2}(\theta_{2}/2)
\label{instdiff}
\eeq
where $\mu_{0}$ is the permeability of free space, $g$ is the g-factor, $\beta$ is the Bohr magneton and $C$ the spin concentration for each hyperfine line. In Figure 1(b) of the main text the plot of 1/\ttwoe\ vs sin$^{2}(\theta/2)$, shows \ttwoe\ can be extended from 190~$\mu$s using the standard Hahn echo sequence to an extrapolated 300~$\mu$s in the limit $\theta_{2}=0$. At $\theta_{2}=0$ the material can be considered to be a homogeneously dilute spin system with no dipolar interaction. Pulse lengths are selected that will fully excite the electron spin transition but will not excite any impurity spins, given as a $\pi/2$ of 100~ns and initial $\pi$ refocusing pulse of 200~ns. Using Eq.~\ref{instdiff} and the slope from Figure 1(b, (green)), the spin concentration, $C$, is calculated as $2.5\times10^{15}$ spins/cm$^{3}$. The spin concentration can then be used to find the dipolar coupling frequency, $\nu_{D}$, between two spins $a$ and $b$ using:
\begin{equation}\label{Hamiltoniandip}
\nu_{D}=\frac{\mu_{0}\mu_{B}^{2}g_{a}g_{b}}{4\pi h r^{3}_{a,b}}(1-3\cos^{2}\theta)
\end{equation}
where $\mu_{0}$ is the permittivity of free-space,  $\mu_{B}$ the Bohr magneton, $g$ the g-factor of the spins, $r_{a,b}$ the distance between the two spins and $\theta$ the angle between the Zeeman field and the inter-spin axis~\cite{sch01dip}. To obtain the mean nearest neighbour distance in a random distribution, $\langle r_{a,b}\rangle$, we can use the expression given by Bhattacharyya and Chakrabarti~\cite{Bha08}. Thus, averaging the angular dependence and considering $S=3/2$, the dipolar coupling frequency at this distance is calculated as 2.5 kHz.

The section below outlines that there is no contribution from \emph{outer} coherences to an electron spin echo of $\tau>70~\mu$s and thus the \ttwoe\ measurement above can be considered due to the \emph{inner} electron coherences, $m_{S}=\pm1/2$, which represent our experimental qubit state.

\vspace{60mm}

\pagebreak
\subsection{Electron `Outer' Coherence: Coherence transfer}

The $m_{S}$ electron spin levels are not sufficiently resolved in frequency that they can be addressed individually by a mw pulse. A mw $\pi/2$ pulse will therefore give a contribution to the, \emph{inner}, $m_{S}=+1/2 \leftrightarrow m_{S}=-1/2$ and \emph{outer} $m_{S}=\pm3/2 \leftrightarrow m_{S}=\pm1/2$ coherences in a standard Hahn echo sequence. A Hahn echo sequence with the addition of resonant rf pulses to remove or impart phase shifts to the different coherences can be used in order to seperate their contribution when measured. This experimental procedure has previously been utilised in `coherence transfer' ENDOR~\cite{sch01CT}. A simple sequence is described by $\pi/2_{mw}-2\pi_{rf}-\pi_{mw}-echo$ (\tab{CT} (A)), whereby the rf 2$\pi$ pulse imparts a $\pi$ phase shift on an electron spin coherence and the signal (in a electron spin-1/2 system) varies from a relative +1 $\rightarrow$ $-$1 or $F_{\rm ENDOR}$ = 1 (where $F_{\rm ENDOR}=1$ is the maximum coherence information available). An electron spin-3/2 system differs from a spin-1/2 system as the rf pulse can affect either the \emph{outer} electron coherences, $m_{I}=\pm3/2$, or both \emph{outer} and \emph{inner} electron coherences, $m_{I}=\pm1/2$. Thus, a series of experiments, shown in ~\tab{CT}, can be envisaged in order to ascertain the contribution from the \emph{outer} coherences to the echo intensity.

\begin{table}[h]
\begin{tabular}[t]{ccc}
{Sequence}&{RF(i)}&{RF(ii)}\\\hline
A&2$\pi_{x}$&0\\
B&$\pi_{x}$&$\pi_{x}$\\
C&$\pi_{x}\pi_{y}$&$\pi_{y}\pi_{x}$\\\hline
\end{tabular}
\caption{Coherence transfer experiments using $\pi/2_{mw}-RF(i)-\pi_{mw}-RF(ii)-echo$. In the table the pulse subscript indicates the axis rotation. In experiments (B) and (C) the splitting of the rf pulse, i.e. an rf $\pi$ pulse before and after the mw refocusing pulse, avoids the Bloch-Siegert shift. In (C) the rf pulses with differing axis of rotation, $\pi_{x}\pi_{y}$, act to induce a $\pi/2$ phase shift on the electron spin echo.}
\label{CT}
\end{table}

\tab{CT2} shows the theoretical results of the  sequences outlined in~\tab{CT} under two scenarios: i) all coherences contribute to the echo observed, and ii) only the \emph{inner} coherence contribute. The experimental results are shown to be entirely consistent with the latter case. This means that there cannot be any contribution from \emph{outer} coherences to the electron spin reference echo used to ascertain the fidelity of the electron-nuclear spin transfer sequence (see main text, $\tau=140~\mu$s).

\begin{table}[h]
\begin{tabular*}{0.8\textwidth}{@{\extracolsep{\fill}}cccccccccccccccc}
&&&\multicolumn{3}{c}{A ($F_{\rm ENDOR}$)}&&&\multicolumn{3}{c}{B ($F_{\rm ENDOR}$)}&&&\multicolumn{3}{c}{C ($F_{\rm ENDOR}$)}\\ \cline{4-6}\cline{9-11}\cline{14-16}
$m_{I}$&&&{all coh.}&{inner only}&{Exp.}&&&{all coh.}&{inner only}&{Exp.}&&&{all coh.}&{inner only}&{Exp.}\\\hline 
$~\pm$1/2&&&0.7 &1&1&&&0.5&0.5&0.5&&&0.7(X):0.3(Y)&1&1\\\
$\pm$3/2&&&0.3&0&0&&&0.3&0&0&&&0.3(X):0.3(Y)&0&0\\\hline 
\end{tabular*}
\caption{Coherence transfer experiments using the sequences outlined in Table I, conducted on the $m_{I}=+3/2$ and $+1/2$ transitions, are compared with two theoretical scenarios: one in which the observed electron spin echo contains maximum contribution from all coherences (`all coh.') and a second in which only the inner coherences contribute to the echo (`inner only'). Brackets indicate the phase of the signal (split between the X and Y channels of the quadrature detector). The experimental results are consistent with the expected contribution if no \emph{outer} coherence is observed on the timescale of 70~$\mu$s.}
\label{CT2}
\end{table}
\vspace{30mm}
\pagebreak
\subsection{Electron `Outer' Coherence: Davies Electron Nuclear Double Resonance (ENDOR)}

The coherence transfer experiments outlined above show that no \emph{outer} electron coherence is present for $\tau$ greater than 70 $\mu$s. Therefore one might na\"ively expect that it is not possible to probe rf transitions in the $m_{S}=\pm3/2$ subspace using Davies ENDOR ($\pi_{mw}-\pi_{rf}-\pi/2_{mw}-\pi_{mw}-echo$~\cite{sch01davies}). In fact it can be shown that such a Davies ENDOR signal can be observed via the electron \emph{inner} coherence. If the reduced $m_{I}=-1/2$ subspace is considered in the electron spin basis [3/2, 1/2, -1/2, -3/2] then an inverted population ($\pi_{mw}$) followed by a $\pi_{rf}$ pulse resonant on the $m_{S}=+3/2$ level will yield the density matrix:

\beq
\left(%
\begin{array}{cccc}
3/2&0&0&0\\
0&-1/2&0&0\\
0&0&1/2&0\\
0&0&0&3/2\\
\end{array}%
\right).
\label{maineq}
\eeq

Completing the sequence with a spin echo readout gives:

\beq
\left(%
\begin{array}{cccc}
3/8&i\sqrt{3}/8&-3\sqrt{3}/8&3i/8\\
i\sqrt{3}/8&9/8&-i/8&-3\sqrt{3}/8\\
-3\sqrt{3}/8&i/8&9/8&i\sqrt{3}/8\\
-3i/8&-3\sqrt{3}/8&-i \sqrt{3}/8&3/8\\
\end{array}%
\right).
\label{maineq2}
\eeq

\begin{figure}[h] \centerline
{\includegraphics[width=5in]{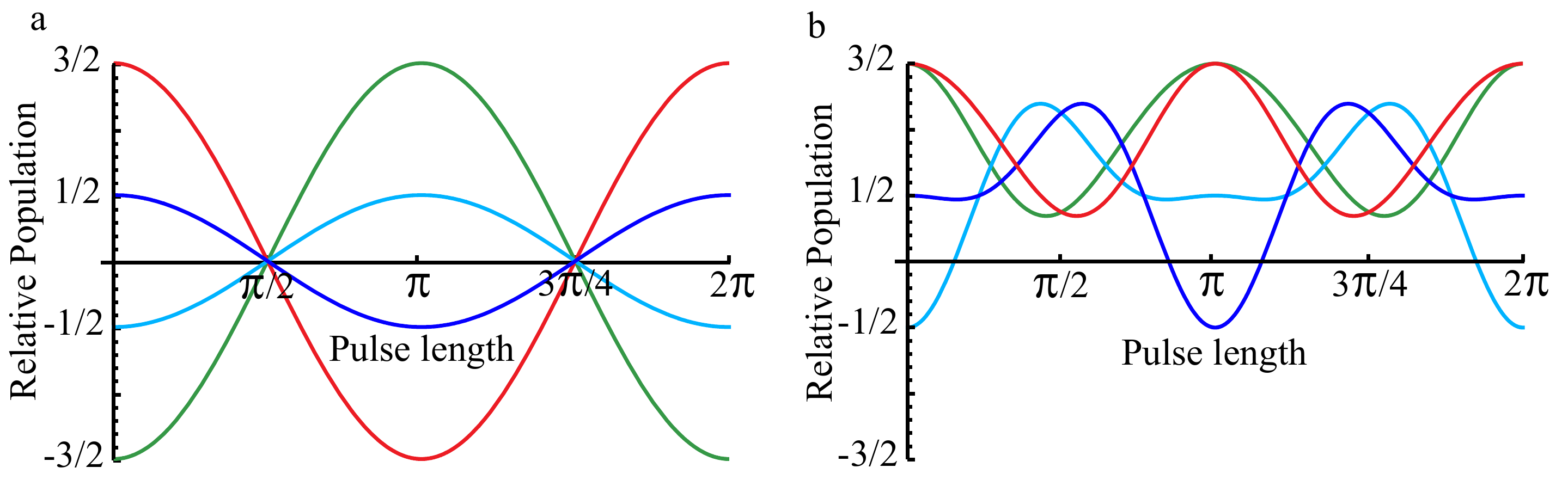}} \caption{Relative population of an electron spin 3/2 system under application of a pulse of varying length a) Starting from an initial $S_z$ or inverted state b) Starting from an inverted state with a $\pi$ rf pulse resonant on the $m_{S}=+3/2$ level (as in a Davies ENDOR experiment on the $m_{S}=+3/2$ line, this state is shown in~\eqn{maineq}).} \label{highspin}
\end{figure}

Observing the \emph{inner} coherence only, through a selective measurement, $F_{\rm ENDOR} = 0.625$. Thus, theoretically one is able to observe the $m_S=\pm3/2$ ENDOR transitions through the \emph{inner} electron coherence. This result can be understood by examining the nature of a $\pi/2$ pulse in the high spin system.  \Fig{highspin} a) shows the relative populations of the high spin system (starting from a thermal state) under the influence of a mw pulse of varying length. As expected, a $\pi/2$ pulse acts to equalise the populations, while a $\pi$ pulse swaps populations. However, in~\Fig{highspin}(b), starting from the state in ~\eqn{maineq} during the davies ENDOR experiment, the subsequent $\pi/2$ pulse acts to separately equalise both the $m_{S}=\pm3/2$ and $m_{S}=\pm1/2$ levels. The $m_{S}=\pm1/2$ levels (in blue) actually increase in relative population on application of a $\pi/2$ pulse relative to the $m_{S}=\pm3/2$ populations. This behaviour gives rise to the observable ENDOR signal which is confirmed experimentally with $F_{\rm ENDOR}\sim0.525$, out to long values of $\tau$ (several \ttwoe). The difference between theoretical and experimental $F_{\rm ENDOR}$ values is due to imperfect pulses. Significantly this means that the full nuclear spin subspace can be exploited even if the \emph{outer} electron coherence times are short (see next section).

\pagebreak
\subsection{Nuclear Coherence in the $m_S=\pm3/2$ subspace}

\begin{figure}[h] \centerline
{\includegraphics[width=3.5in]{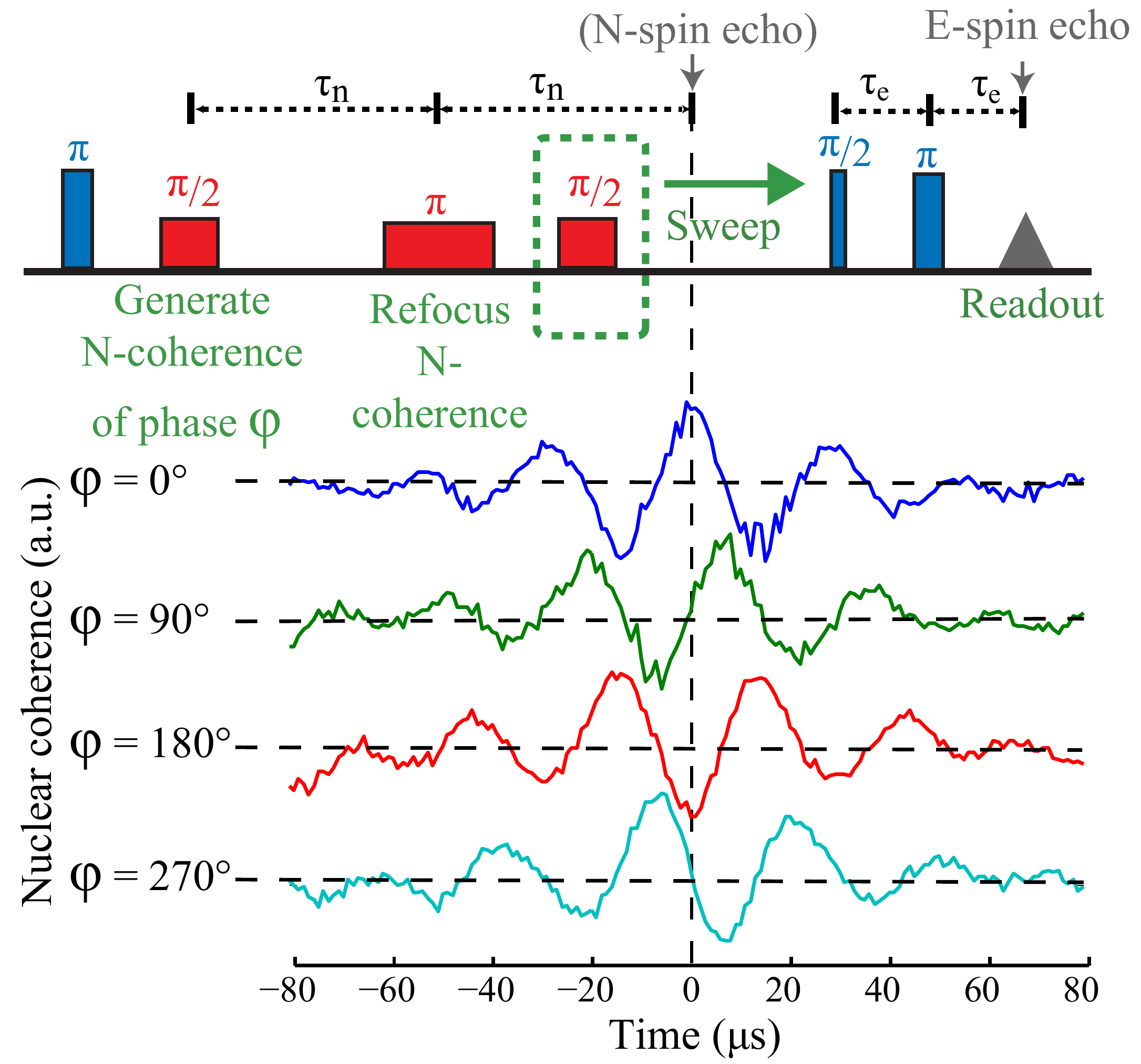}} \caption{Sequence to probe the nuclear coherence for $m_S=\pm3/2$, as described in the text. The initial phase generated in the nuclear spin is faithfully recovered by polarisation transfer and readout using the electron spin. The nuclear coherence is off-resonance in order to more clearly observe the recovered phase.} \label{outer}
\end{figure}

The main text discussed the transfer of qubit states from the electron spin \emph{inner} levels to the nuclear spin $m_{S}=\pm1/2$ manifolds. Transfer from of an electron spin \emph{outer} coherence to the nuclear spin is not possible, but the nuclear spin transitions in the $m_{S}=\pm3/2$ manifolds could be still used to store quantum information and make greater use of the multi-levelled system. A sequence can be applied to probe directly \ttwon\ of the $m_{S}=\pm3/2$ manifolds not possible in the low concentration limit using NMR. The sequence (\Fig{outer}) uses an rf $\pi$/2 pulse on the $m_{S}=+3/2$ manifold to directly generate a nuclear coherence of arbitrary phase (given by $\varphi$). The coherence is then stored for a time ($2\tau_{n}$) and transferred to a nuclear polarisation via a second rf $\pi$/2 pulse for readout by a selective electron spin echo. The rf frequency in \Fig{outer} is set to 40~kHz off-resonance and the second $\pi$/2 pulse swept to show the fringe pattern of the nuclear coherence, read via the electron spin. The phase varies as expected with the initial coherence generated. The \ttwon\ can then be measured with the rf on resonance and the electron spin readout in fixed position, incrementing  $\tau_{n}$. A subtlety of this experiment is that it only allows measurement of \ttwon\ in the region where \tonee~$>$~\ttwon, as it relies on the polarisation given by the electron spin inversion (inital $\pi$ mw pulse). Remarkably, as the measurement is effectively of electron spins that have not relaxed, when \tonee~$\le$~\ttwon\ the nuclear spin shows no decoherence, an observation that can be confirmed by modelling of the system. In the low temperature region we find \tonee~$>$~\ttwon\ and the extracted nuclear coherence times for the $m_{S}=\pm3/2$ are found to be of similar order to those in the $m_{S}=\pm1/2$ manifolds. This indicates a a nuclear decoherence mechanism which acts similarly for all nuclear spin manifolds.
\vspace{30mm}
\pagebreak
\subsection{Nuclear Phase Gates}

To confirm the robustness of the transfer sequence we can apply a specific time-varying phase to the nuclear spin state which can then be verified in the measurement to show that the qubit must have been held entirely in the nuclear spin state. To apply this phase we employ a geometric phase gate similar to that of Aharonov-Anandan gate~\cite{aha87, pines88}, consisting of two resonant $\pi$ rf pulses which will be invariant to spin populations. Varying the phase of the second rf pulse relative to the first by $\delta \phi$, the nuclear spin qubit undergoes a path around the Bloch sphere, resulting in a geometric phase. In the basis $(m_S,m_I)=[(1/2,1/2),(-1/2,1/2),(1/2,-1/2),(1/2,-1/2)]$ a $\pi_{0}$ followed by a $\pi_{\phi}$ pulse (where subscript indicates the phase) applies a phase of 2$\phi$ to the nuclear coherence, given by the operator:
\beq
U(\phi)= 
\left(%
\begin{array}{cccc}
e^{i\phi}& 0 & 0 & 0\\
0& 1 & 0 & 0\\
0& 0& e^{i\phi}& 0\\
0& 0 & 0& 1 
\end{array}%
\right)
\label{phase}
\eeq

\begin{figure}[h] \centerline
{\includegraphics[width=5in]{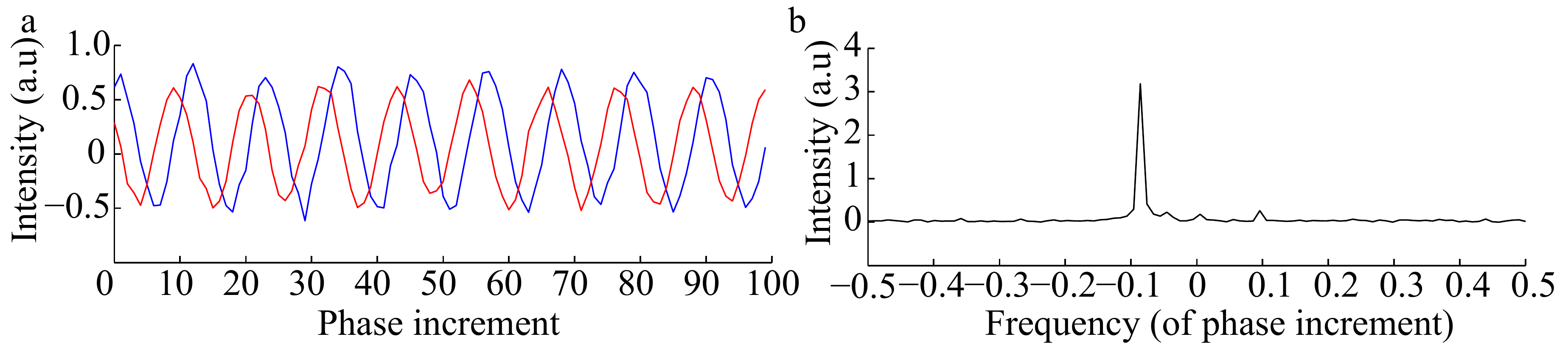}} \caption{(a) Incrementing the phase of a nuclear coherence (b) Fourier transform of the phase increment with a peak corresponding to a nuclear coherence, in this case set to -2$\phi$. Additional frequencies are due pulse imperfections} \label{fourier}
\end{figure}

\begin{figure}[h] \centerline
{\includegraphics[width=3in]{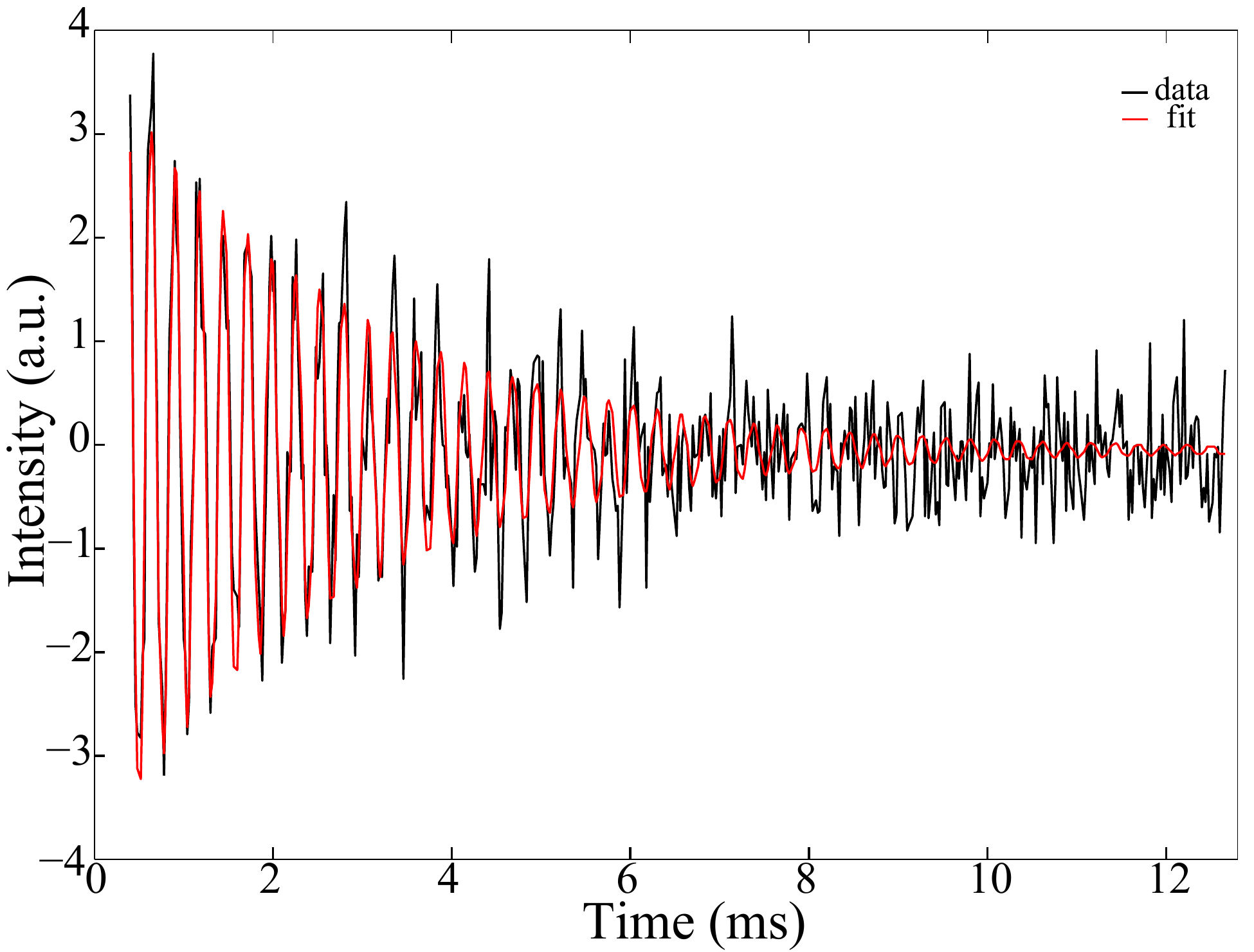}} \caption{Nuclear decoherence curve at 60~K with phase increment (512 points) on the nuclear transition using a geometric phase gate. The data is fit to damped oscillatory function with an increment frequency 2$\phi$ as expected given a phase increment of $\phi$, according to the operator in~\eqn{phase}} \label{nucphase}
\end{figure}

Experimentally this is implemented by applying the two rf pulses immediately after the refocusing rf pulse (main text, Figure 1(d)) and incrementing the phase of the second using a using a Rohde and Schwarz AFQ 100B. A Fourier transform of the phase-incremented signal, as shown in \Fig{fourier}, shows the frequency component corresponding to signal from a nuclear coherence only, given in this case by -2$\phi$. A \ttwon\ measurement can be conducted whilst incrementing the phase (such that the x axis gives both decoherence time and phase increment (per point)) to show that this coherence contributes fully to the decaying signal. The oscillating experimental data, \Fig{nucphase}, fits well to a damped oscillatory function, $e^{-t/T_{2}}\cos(\omega t)$, which at 60~K gives $T_{2n}=2.75 \pm $0.27 and an increment frequency of 0.0891(1). This is entirely consistent with the $T_{2n}$ given by the standard measurement and is at the expected nuclear coherence frequency (2$\phi$).

\pagebreak
\section{Transfer Fidelity: Store and Re-store sequence}

\begin{figure*}[h] \centerline
{\includegraphics[width=7in]{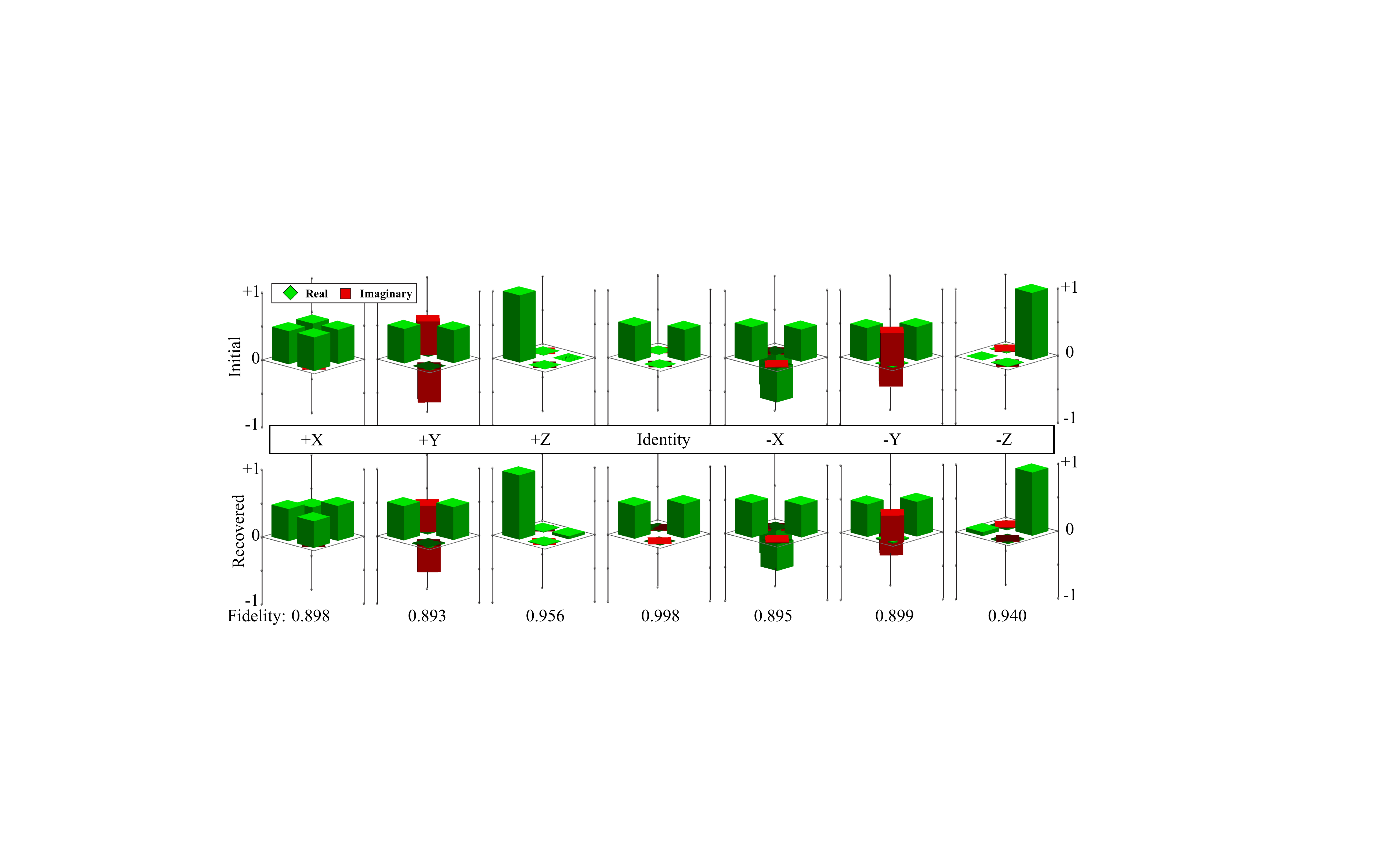}} \caption{Density matrix tomography of the nominal +X, +Y, +Z, -X, -Y, -Z and Identity pseudopure states for the initial state and that recovered after storage in the nuclear spin degree of freedom. The fidelity is calculated using $F$=$\langle \psi |  \rho_{1} | \psi \rangle$, where  $\rho_{0} =  | \psi \rangle \langle \psi | $.} \label{fidelityall}
\end{figure*}

Arbitrary qubit states can be transferred between electron and nuclear degrees of freedom as described in the main text.  The measured transfer fidelities, $F$, are shown in \Fig{fidelityall} and compare the recovered echo from the transfer sequence ($\rho_{1}$) with an ordinary Hahn echo ($\rho_{0}$) with the same electron dephasing time ($\tau=\tau_{e1}+\tau_{e2}$). We use the convention $F=\langle \psi |  \rho_{1} | \psi \rangle$, where  $\rho_{0} =  | \psi \rangle \langle \psi | $. The initial states are prepared by: varying the phase of the initial $\pi$/2 mw pulse for $\pm$X and $\pm$Y; application of a $\pi$ pulse for $+$Z; a $\pi$ pulse followed by a wait time of ln(2)\tonee\ for the Identity and no initial pulse for the $-$Z state. These states can then be used to perform quantum process tomography to produce the process matrix, $\chi$, as shown in the main text.

\begin{figure}[h] \centerline
{\includegraphics[width=7in]{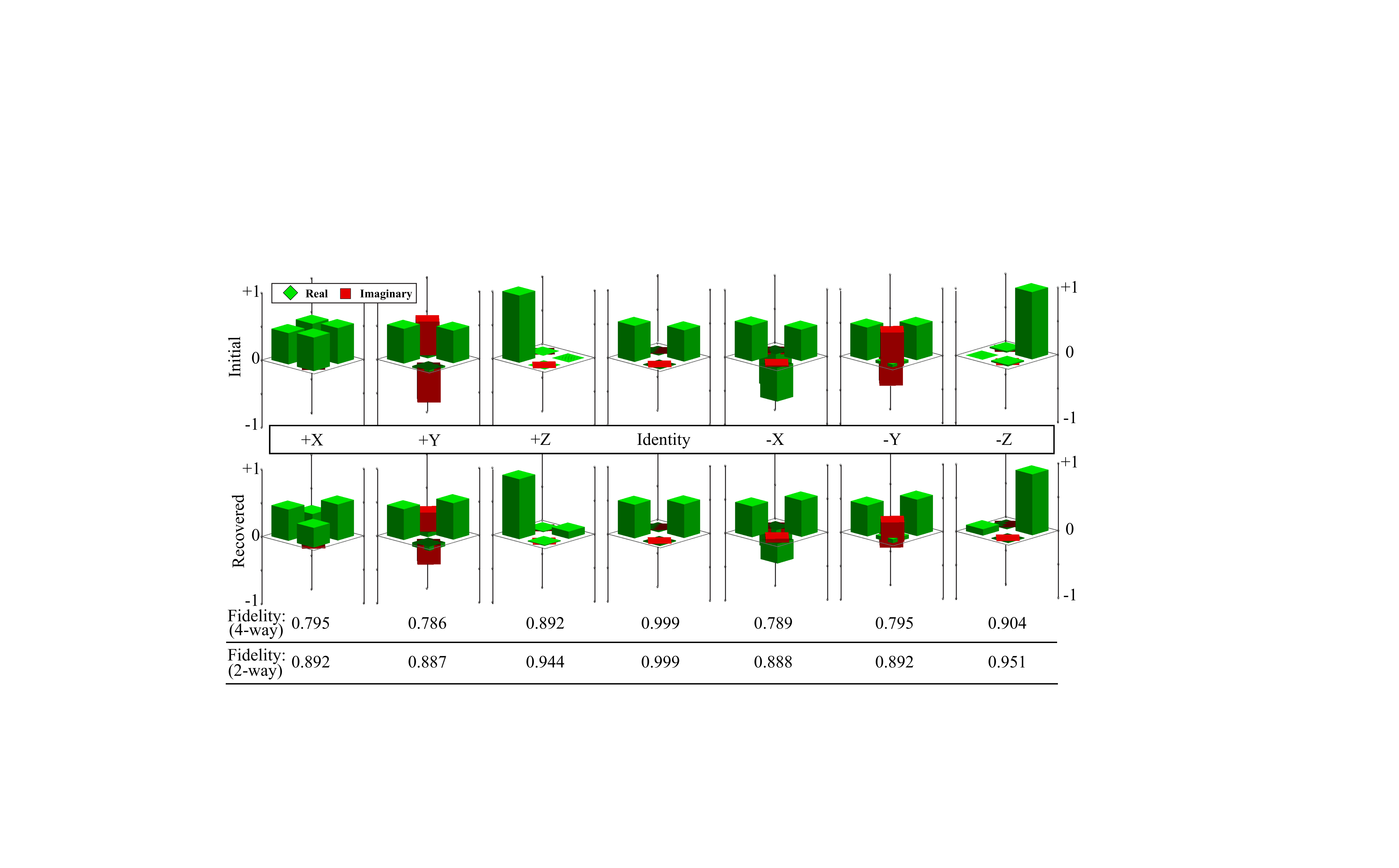}} \caption{{ `Re-store' fidelity. Density matrix tomography of the nominal +X, +Y, +Z, -X, -Y, -Z and Identity pseudopure states for the initial state and that recovered after storage twice in the nuclear spin degree of freedom. The 4-way fidelity is calculated using $F=\langle \psi |  \rho_{1} | \psi \rangle$, where  $\rho_{0} =  | \psi \rangle \langle \psi | $. The 2-way fidelity is given by the square route of the 4-way fidelity.}} \label{restore}
\end{figure}

The robustness of the transfer sequence can also be examined by `storing' the qubit state in the nuclear spin, returning it to the electron spin and then `re-storing' it in the nuclear spin before readout in the electron spin. Such a sequence will give a 4-way transfer fidelity (expected to be the square of the 2-way fidelity) and allow the storage of the \emph{inner} electron coherence ($m_{I}=-1/2, M_{S}=\pm1/2$ levels) only, as the initial storage acts to remove the \emph{outer} unwanted electron coherences. The restore fidelity is given in \Fig{restore} in agreement with the standard transfer sequence (\Fig{fidelityall}).\\

\subsection{Transfer Fidelity: BB1}

\begin{figure}[h] \centerline
{\includegraphics[width=3.5in]{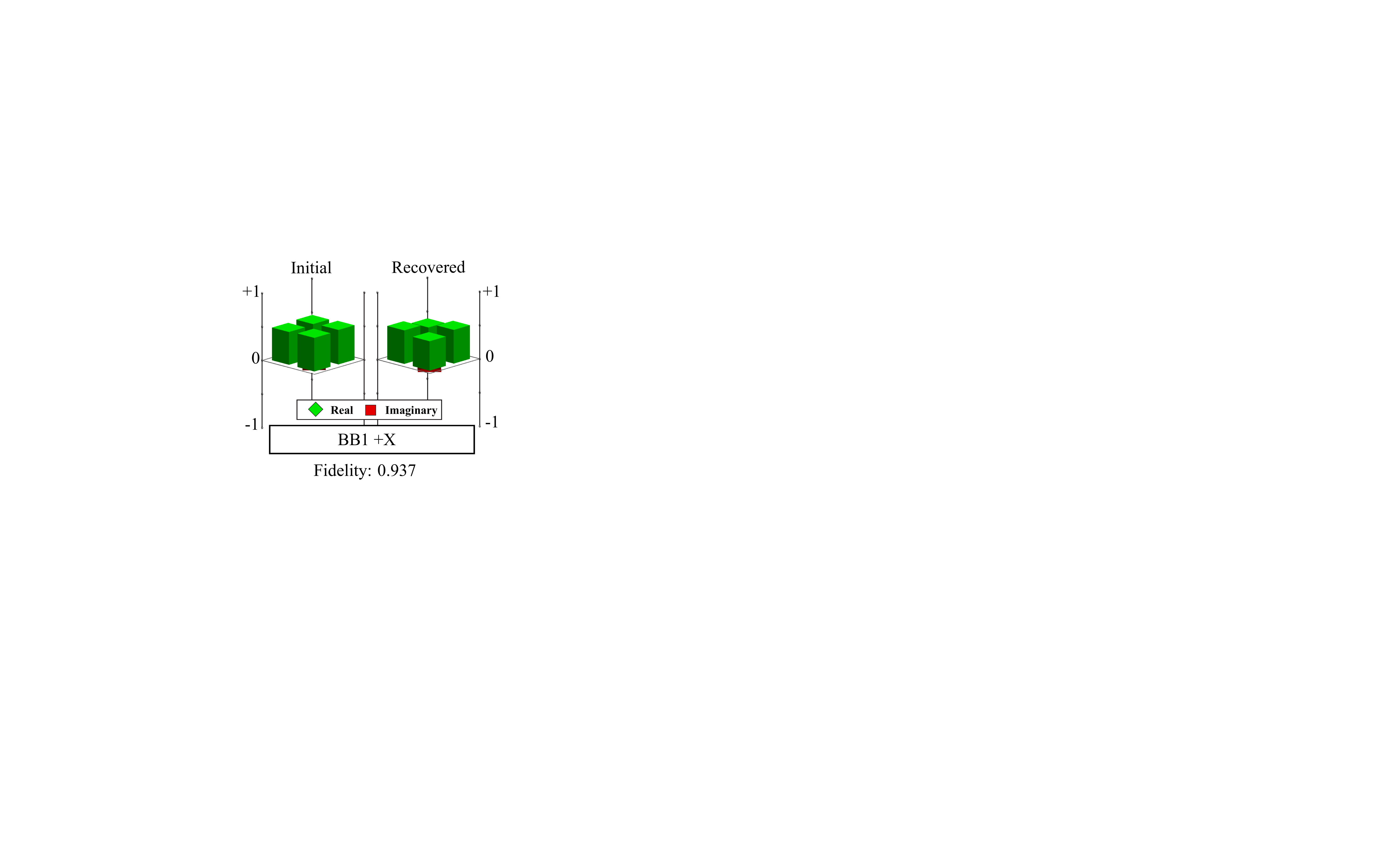}} \caption{When each of the microwave pulses used is replaced with an error-correcting (BB1) pulse, the two-way transfer fidelity is improved substantially, shown here for the case of +X.} \label{BB1}
\end{figure}

The fidelity of the sequence can be improved by replacing each of the microwave pulses in the transfer sequence (main text, Figure 1 (d)) with an error correcting microwave pulse~\cite{NMRBB1, EPRBB1}.  A $\pi$ pulse is optimised with the error correcting sequence $\pi$(0$^{\circ}$)-$\pi$(104.5$^{\circ}$)-2$\pi$(313.4$^{\circ}$)-$\pi$(104.5$^{\circ}$), where brackets indicate the axis of rotation in degrees. Applying this to both the reference and full transfer sequence for the +X state, a fidelity of 94$\%$ is achieved, as shown in ~\Fig{BB1}.

\end{document}